# Precoding for the AWGN Channel with Discrete Interference

Hamid Farmanbar and Amir K. Khandani
Coding and Signal Transmission Laboratory
Department of Electrical and Computer Engineering
University of Waterloo
Waterloo, Ontario, N2L 3G1, Canada
Email: {hamid,khandani}@cst.uwaterloo.ca

**Abstract**

$M$-ary signal transmission over AWGN channel with additive $Q$-ary interference where the sequence of i.i.d. interference symbols is known causally at the transmitter is considered. Shannon's theorem for channels with side information at the transmitter is used to formulate the capacity of the channel. It is shown that by using at most $MQ - Q + 1$ out of $M^Q$ input symbols of the *associated* channel, the capacity is achievable. For the special case where the Gaussian noise power is zero, a sufficient condition, which is independent of interference, is given for the capacity to be $\log_2 M$ bits per channel use. The problem of maximization of the transmission rate under the constraint that the channel input given any current interference symbol is uniformly distributed over the channel input alphabet is investigated. For this setting, the general structure of a communication system with optimal precoding is proposed. The extension of the proposed precoding scheme to continuous channel input alphabet is also investigated.

**Index Terms**

Causal side information, interference, channel capacity, precoding, linear programming, integer programming.

# I. INTRODUCTION

Information transmission over channels with known interference at the transmitter has been a major focus of research due to its application in various communication problems. A remarkable result on such channels was obtained by Costa who showed that the capacity of the additive white Gaussian noise (AWGN) channel with additive Gaussian i.i.d. interference, where the sequence of interference symbols is known non-causally at the transmitter, is the same as the capacity of AWGN channel [1]. Therefore, the interference does not incur any loss in the capacity. This result was extended to arbitrary interference (random or deterministic) Erez *et al.* [2]. Following Costa's "Writing on dirty paper" famous title [1], coding strategies for the channel with non-causally known interference at the transmitter are referred to as "dirty paper coding" (DPC).

Transmission over multiple-input multiple-output (MIMO) broadcast channel is an important application of DPC. In such systems, for a given user, the signals sent to the other users are considered as interference. Since all signals are known to the transmitter, dirty paper coding can be used after some linear preprocessing [3]. It was shown that DPC in fact achieves the sum capacity of the MIMO broadcast channel [4], [5], [6]. Most recently, it has been shown that the same is true for the entire capacity region of the MIMO broadcast channel [7]. Another important application of DPC is information embedding or watermarking [8], [9], [10], where a host signal is modeled as interference onto which a watermark signal is embedded.

The result obtained by Costa does not hold for the case that the sequence of interference symbols is known causally at the transmitter. In fact, the capacity is unknown in this case and unlike the non-causal knowledge setting, the capacity depends on the interference. The only definitive result in this case is due to Erez *et al.* [2] who showed that, for the worst-case interference, at the limit of high SNR, the loss in capacity due to not having the future samples of the interference at the transmitter is exactly the ultimate shaping gain $\frac{1}{2}\log\left(\frac{2\pi e}{12}\right) \approx 0.254$ bit.

In this paper, we consider the AWGN channel with i.i.d. additive discrete interference where the sequence of interference symbols is known causally at the transmitter. The discrete interference model is more appropriate for many practical applications. For example, in the MIMO broadcast channel, due to the fact that in practice the user signals are chosen from finite constellations, the interference caused by the other users is discrete rather than continuous. We are interested in both capacity of the channel and precoding schemes for the channel.

The rest of the paper is organized as follows. In section II, we provide some background on channels with side information at the encoder. In section III, we introduce our channel model. In section IV, we investigate the capacity of the channel. In section V, we consider maximizing the transmission rate under the constraint that the channel input given any current interference symbol is uniformly distributed over the channel input alphabet. The general structure of a communication system for the channel with causally-known discrete interference is given in section VI. We extend the uniform transmission scheme to continuous-input alphabet in section VII. We conclude this paper in section VIII.

## II. Channels with Side Information at the Transmitter

Channels with known interference at the transmitter are special case of channels with side information at the transmitter which were considered first by Shannon [11].

Shannon considered a discrete memoryless channel (DMC) whose transition matrix depends on the channel state. A state-dependent discrete memoryless channel (SD-DMC) is defined by a finite input alphabet $\mathcal{X}$, a finite output alphabet $\mathcal{Y}$, and transition probabilities $p(y|x, s)$, where the state $s$ takes on values in a finite alphabet $\mathcal{S}$. The block diagram of a state-dependent channel with state information at the encoder is shown in fig. 1.

We may consider two settings for the knowledge of state sequence at the encoder: causal or non-causal. In the causal knowledge setting, the encoder maps a message $w$ into $\mathcal{X}^n$ such that the channel input at time $i$ is a function of the message $w$ and the state sequence up to the time $i$, $i = 1, 2, \ldots, n$, whereas in the non-causal knowledge setting, the encoder observes the entire state sequence to generate every symbol of the code sequence.

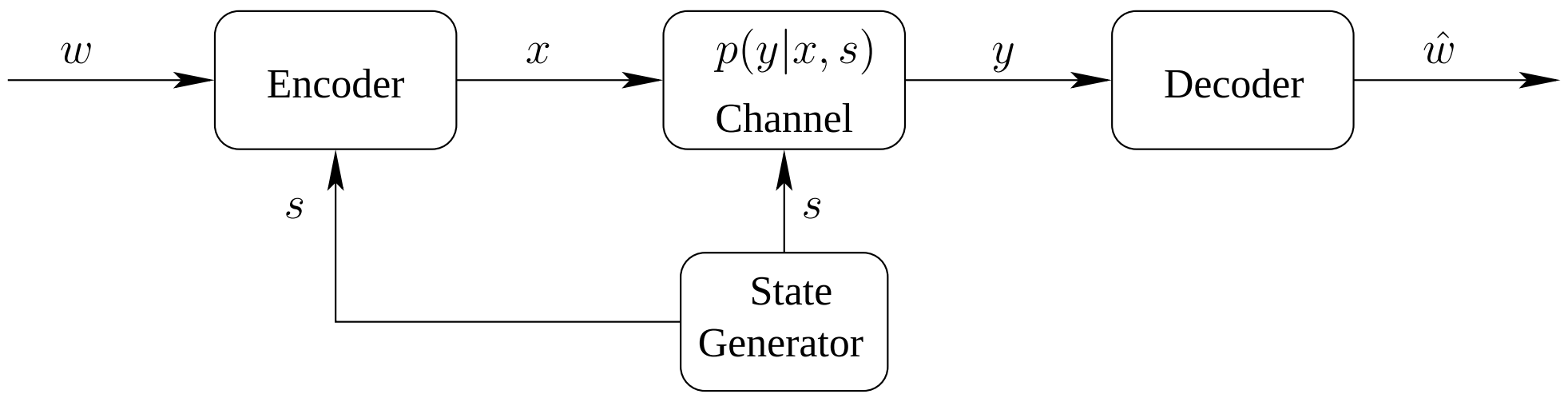

Fig. 1. SD-DMC with state information at the encoder.

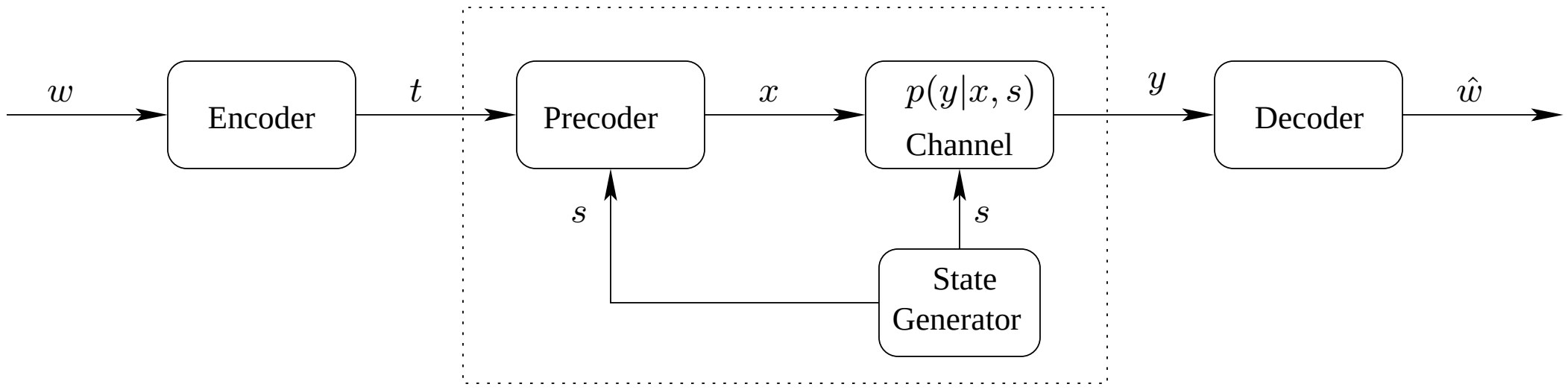

Fig. 2. The associated regular DMC.

Shannon considered the case where the i.i.d. state sequence is known causally at the encoder and obtained the capacity formula [11]. The case where the i.i.d. state sequence is known non-causally at the encoder was considered by Kuznetsov and Tsybakov in the context of coding for memories with defective cells [12]. Gel'fand and Pinsker obtained the capacity formula for this case [13].

Shannon's capacity formula was generalized by Salehi [14] for the case that a noisy version of the state sequence is available at both encoder and decoder. Caire and Shamai [15] investigated the case that the state sequence is not memoryless. The capacity results with non-causal side information at the encoder were generalized to the case were rate-limited side information is available at both encoder and decoder [16], [17].

Shannon [11] showed that the capacity of an SD-DMC where the i.i.d. state sequence is known causally at the encoder is equal to the capacity of an *associated* regular (without state) DMC with an extended input alphabet $\mathcal{T}$ and the same output alphabet $\mathcal{Y}$. The input alphabet of the associated channel is the set of all functions from the state alphabet to the input alphabet of the state-dependent channel. There are a total of $|\mathcal{X}|^{|\mathcal{S}|}$ of such functions, where $|.|$ denotes the cardinality of a set. Any of the functions can be represented by a $|\mathcal{S}|$-tuple $(x_1, x_2, \ldots, x_{|\mathcal{S}|})$ of elements of $\mathcal{X}$, implying that the value of the function at state $s$ is $x_s, s = 1, 2, \ldots, |\mathcal{S}|$.

The transition probabilities for the associated channel are given by [11]

$$p(y|t) = \sum_{s=1}^{|\mathcal{S}|} p(s)p(y|x_s, s), \tag{1}$$

where $t$ denotes the the function represented by $(x_1, x_2, \ldots, x_{|\mathcal{S}|})$. Also,

$$p(y(1)\cdots y(n)|t(1)\cdots t(n)) = \prod_{i=1}^{n} p(y(i)|t(i)), \tag{2}$$

where $i$ denotes the time index. The capacity is given by [11]

$$C = \max_{p(t)} I(T;Y), \tag{3}$$

where the maximization is taken over the probability mass function (pmf) of the random variable $T$.

Any encoding and decoding scheme for the associated channel can be translated into an encoding and decoding scheme for the original state-dependent channel with the same probability of error [11]. An encoder for the associated channel encodes a message $w$ to $(t(1), \ldots, t(n))$. The translated encoding scheme for the original state-dependent channel is to map the message $w$ to $(x(1), x(2), \ldots, x(n))$, where $x(i) = s$th component of $t(i)$ if the state at time $i$ is $s$, $s = 1, 2, \ldots, |\mathcal{S}|$, and $i = 1, 2, \ldots, n$. The block diagram of the associated regular DMC is shown in fig. 2.

In the capacity formula (3), we can alternatively replace the random variable $T$ with $(X_1, \ldots, X_{|\mathcal{S}|})$, where $X_s$ is the random variable that represents the input to the state-dependent channel when the state is $s, s = 1, \ldots, |\mathcal{S}|$.

## III. The Channel Model

We consider data transmission over the channel

$$Y = X + S + N, \tag{4}$$

where $X$ is the channel input, which takes on values in a fixed real constellation

$$\mathcal{X} = \{x_1, x_2, \ldots, x_M\}, \tag{5}$$

$Y$ is the channel output, $N$ is additive white Gaussian noise with power $P_N$, and the interference $S$ is a discrete random variable that takes on values in

$$\mathcal{S} = \{s_1, s_2, \ldots, s_Q\} \tag{6}$$

with probabilities $r_1, r_2, \ldots, r_Q$, respectively. The sequence of i.i.d. interference symbols is known causally at the encoder.

The above channel can be considered as a special case of state-dependent channels considered by Shannon with one exception, that the channel output alphabet is continuous. In our case, the likelihood function $f_{Y|X,S}(y|x,s)$ is used instead of the transition probabilities. We denote the input to the associated channel by $T$, which can also be represented as $(X_1, X_2, \ldots, X_Q)$, where $X_j$ is the random variable that represents the channel input when the current interference symbol is $s_j$, $j = 1, \ldots, Q$.

The likelihood function for the associated channel is given by

$$\begin{aligned} f_{Y|T}(y|t) &= \sum_{j=1}^{Q} r_j f_{Y|X,S}(y|x_{i_j}, s_j) \\ &= \sum_{j=1}^{Q} r_j f_N(y - x_{i_j} - s_j), \end{aligned} \tag{7}$$

where $f_N$ denotes the pdf of the Gaussian noise $N$, and $t$ is the input symbol of the associated channel represented by $(x_{i_1}, x_{i_2}, \ldots, x_{i_Q})$. The pdf of $Y$ is then given by

$$
\begin{aligned}
f_Y(y) &= \sum_{i_1=1}^{M} \cdots \sum_{i_Q=1}^{M} p_{i_1 i_2 \cdots i_Q} \left( \sum_{j=1}^{Q} r_j f_N(y - x_{i_j} - s_j) \right) \\
&= \sum_{j=1}^{Q} r_j \sum_{i=1}^{M} p_i^{(j)} f_N(y - x_i - s_j), \qquad (8)
\end{aligned}
$$

where $p_{i_1 i_2 \cdots i_Q} = \Pr\{X_1 = x_{i_1}, \ldots, X_Q = x_{i_Q}\}$, $p_i^{(j)} = \Pr\{X_j = x_i\}$.

## IV. The capacity

The capacity of the associated channel, which is the same as the capacity of the original state-dependent channel, is the maximum of $I(T;Y) = I(X_1 X_2 \cdots X_Q; Y)$ over the joint pmf values $p_{i_1 i_2 \cdots i_Q}$, i.e.,

$$
C = \max_{p_{i_1 i_2 \cdots i_Q}} I(X_1 X_2 \cdots X_Q; Y). \qquad (9)
$$

The mutual information between $T$ and $Y$ is the difference between differential entropies $h(Y)$ and $h(Y|T)$. It can be seen from (8) that $f_Y(y)$, and hence $h(Y)$, are uniquely determined by the marginal pmfs $\{p_i^{(j)}\}_{i=1}^{M}$, $j = 1, \ldots, Q$. The conditional entropy $h(Y|T)$ is given by

$$
\begin{aligned}
h(Y|T) &= h(Y|X_1 X_2 \cdots X_Q) \\
&= \sum_{i_1=1}^{M} \cdots \sum_{i_Q=1}^{M} p_{i_1 \cdots i_Q} h(Y|X_1 = x_{i_1}, \ldots, X_Q = x_{i_Q}) \\
&= \sum_{i_1=1}^{M} \cdots \sum_{i_Q=1}^{M} p_{i_1 \cdots i_Q} h_{i_1 \cdots i_Q}, \qquad (10)
\end{aligned}
$$

where $h_{i_1 \cdots i_Q} = h(Y|X_1 = x_{i_1}, \ldots, X_Q = x_{i_Q})$.

There are $M^Q$ variables involved in the maximization problem (9). Each variable represents the probability of an input symbol of the associated channel. The following theorem regards the number of nonzero variables required to achieve the maximum in (9).

*Theorem 1:* The capacity of the associated regular channel is achieved by using at most $MQ-Q+1$ out of $M^Q$ inputs with nonzero probabilities.

*Proof:* Denote by $\{\hat{p}_i^{(j)}\}_{i=1}^M$ the pmf of $X_j$, $j=1,2,\ldots,Q$, induced by a capacity-achieving joint pmf $\{\hat{p}_{i_1\cdots i_Q}\}_{i_1,\ldots,i_Q=1}^M$. We limit the search for a capacity-achieving joint pmf to the set of joint pmfs that yield the same marginal pmfs as $\{\hat{p}_{i_1\cdots i_Q}\}_{i_1,\ldots,i_Q=1}^M$. By limiting the search to this set, the maximum of $I(X_1\cdots X_Q;Y)$ remains unchanged (since the capacity-achieving joint pmf $\{\hat{p}_{i_1\cdots i_Q}\}_{i_1,\ldots,i_Q=1}^M$ is in the new set). But all joint pmfs in the new set yield the same $h(Y)$ since they induce the same marginal pmfs on $X_1,\ldots,X_Q$. Therefore, the maximization problem in (9) reduces to the linear minimization problem

$$\begin{aligned}
\min_{p_{i_1\cdots i_Q}} \quad & \sum_{i_1=1}^{M}\cdots\sum_{i_Q=1}^{M} h_{i_1\cdots i_Q}p_{i_1\cdots i_Q} \\
\text{subject to} \quad & \\
& \sum_{i_2=1}^{M}\cdots\sum_{i_Q=1}^{M} p_{i_1\cdots i_Q} = \hat{p}_{i_1}^{(1)}, \qquad i_1 = 1,2,\ldots,M, \\
& \qquad\vdots \qquad\qquad\qquad\qquad\qquad\qquad \vdots \\
& \sum_{i_1=1}^{M}\cdots\sum_{i_{Q-1}=1}^{M} p_{i_1\cdots i_Q} = \hat{p}_{i_Q}^{(Q)}, \qquad i_Q = 1,2,\ldots,M, \\
& p_{i_1\cdots i_Q} \geq 0, \qquad\qquad\qquad i_1,\ldots,i_Q = 1,2,\ldots,M.
\end{aligned} \tag{11}$$

There are $MQ$ equality constraints in (11) out of which $MQ-Q+1$ are linearly independent. From the theory of linear programming, the minimum of (11), and hence the maximum of $I(X_1\cdots X_Q;Y)$, is achieved by a feasible solution with at most $MQ-Q+1$ nonzero variables. ■

Theorem 1 states that at most $MQ-Q+1$ out of $M^Q$ inputs of the associated channel are needed to be used with positive probability to achieve the capacity. However, in general, one does not know which of the inputs must be used to achieve the capacity. If we knew the marginal pmfs for $X_1,\ldots,X_Q$ induced by a capacity-achieving joint pmf, we could obtain the capacity-achieving joint pmf itself by solving the linear program (11).

*A. The Noise-Free Channel*

We consider a special case where the noise power is zero in (4). In the absence of noise, the channel output $Y$ takes on at most $MQ$ different values since different $X$ and $S$ pairs may yield the same sum. If $Y$ takes on exactly $MQ$ different values, then it is easy to see that the capacity is $\log_2 M$ bits [1]: The decoder just needs to partition the set of all possible channel output values into $M$ subsets of size $Q$ corresponding to $M$ possible inputs, and decide that which subset the current received symbol belongs to.

In general, where the cardinality of the channel output symbols can be less than $MQ$, we will show that under some condition on the channel input alphabet, there exists a coding scheme that achieves the rate $\log_2 M$ in one use of the channel. We do this by considering a one-shot coding scheme which uses only $M$ (out of $M^Q$) inputs of the associated channel.

In a one-shot coding scheme, a message is encoded to a single input of the associated channel. Any input of the associated channel can be represented by a $Q$-tuple composed of elements of $\mathcal{X}$. Given that the current interference symbol is $s_j$, the $j$th element of the $Q$-tuple is sent through the channel. Therefore, one single message can result in (up to) $Q$ symbols at the output. For convenience, we consider the output symbols corresponding to a single message as a multi-set[2] of size (exactly) $Q$. If the $M$ multi-sets at the output corresponding to $M$ different messages are mutually disjoint, reliable transmission through the channel is possible.

Unfortunately, we cannot always find $M$ inputs of the associated channel such that the corresponding multi-sets are mutually disjoint. For example, consider a channel with the input alphabet $\mathcal{X} = \{0, 1, 2, 4\}$ and the interference alphabet $\mathcal{S} = \{0, 1, 3\}$. It is easy to check that for this channel we cannot find four triples composed of elements of $\mathcal{X}$ such that the corresponding multi-sets are mutually disjoint. In fact, by entropy calculations,

[1] This is true even if the interference sequence is unknown to the encoder.

[2] A multi-set differs from a set in that each member may have a multiplicity greater than one. For example, $\{1, 3, 3, 7\}$ is a multi-set of size four where 3 has multiplicity two.

we can show that the capacity of the channel in this example is less than $2$ bits.

However, if we impose some constraint on the channel input alphabet, the rate $\log_2 M$ is achievable.

*Theorem 2:* Suppose that the elements of the channel input alphabet $\mathcal{X}$ form an arithmetic progression. Then the capacity of the noise-free channel

$$Y = X + S, \tag{12}$$

where the sequence of interference symbols is known causally at the encoder equals $\log_2 M$ bits.

*Proof:* Let $\mathcal{Y}^{(q)}$ be the set of all possible outputs of the noise-free channel when the interference symbol is $s_q$, i.e.,

$$\mathcal{Y}^{(q)} = \{x_1 + s_q, x_2 + s_q, \ldots, x_M + s_q\}, \quad q = 1, \ldots, Q. \tag{13}$$

The union of $\mathcal{Y}^{(q)}$s is the set of all possible outputs of the noise-free channel.

Without loss of generality, we can assume that $s_1 < s_2 < \cdots < s_Q$. The elements of $\mathcal{Y}^{(q)}$ form an arithmetic progression, $q = 1, \ldots, Q$. Furthermore, these $Q$ arithmetic progressions are shifted versions of each other.

We prove by induction on $Q$ that there exist $M$ mutually-disjoint multi-sets of size $Q$ composed of the elements of $\mathcal{Y}^{(1)}, \mathcal{Y}^{(2)}, \ldots, \mathcal{Y}^{(Q)}$ (one element from each). If we can find such $M$ multi-sets of size $Q$, then we can obtain the corresponding $M$ $Q$-tuples of elements of $\mathcal{X}$ by subtracting the corresponding interference terms from the elements of the multi-sets. These $M$ $Q$-tuples can serve as the inputs of the associated channel to be used for sending any of $M$ distinct messages through the channel without error in one use of the channel, hence achieving the rate $\log_2 M$ bits per channel use.

For $Q = 1$, the statement of the theorem is true since we can take $\{x_1 + s_1\}, \{x_2 + s_1\}, \ldots, \{x_M + s_1\}$ as mutually-disjoint sets of size one.

Assume that there exist $M$ mutually-disjoint multi-sets of size $Q = q$. For $Q = q+1$, we will have the new set of channel outputs $\mathcal{Y}^{(q+1)} = \{x_1+s_{q+1}, x_2+s_{q+1}, \ldots, x_M+s_{q+1}\}$. We consider two possible cases:

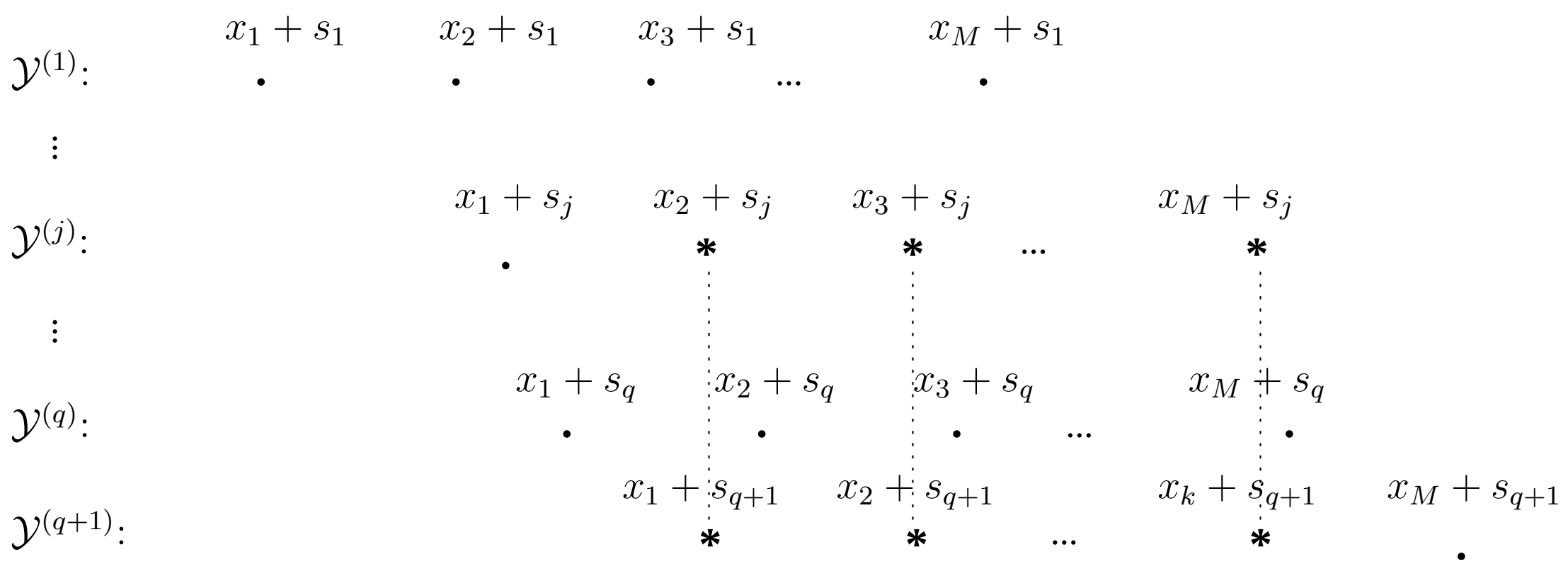

Fig. 3. The elements of $\mathcal{Y}^{(1)},\ldots,\mathcal{Y}^{(q+1)}$ shown as shifted version of each other. The elements of $\mathcal{Y}^{(q+1)}$ up to $x_k+s_{q+1}$ appear in $\mathcal{Y}^{(j)}$.

*Case 1*: None of the elements of $\mathcal{Y}^{(q+1)}$ appear in any of the multi-sets of size $Q=q$.

In this case, we include the elements of $\mathcal{Y}^{(q+1)}$ in the $M$ multi-sets arbitrarily (one element is included in each multi-set). It is obvious that the resulting multi-sets of size $Q=q+1$ are mutually disjoint.

*Case 2*: Some of the elements of $\mathcal{Y}^{(q+1)}$ appear in some of the multi-sets of size $Q=q$.

Suppose that the largest element of $\mathcal{Y}^{(q+1)}$ which appears in any of the sets $\mathcal{Y}^{(1)},\ldots,$ $\mathcal{Y}^{(q)}$ (or equivalently, in any of the multi-sets of size $Q=q$) is $x_k+s_{q+1}$ for some $1\le k\le M-1$. Then since $\mathcal{Y}^{(q+1)}$ is shifted version of each $\mathcal{Y}^{(1)},\ldots,\mathcal{Y}^{(q)}$ and $s_{q+1}>s_q>\cdots>s_1$, exactly one of the sets $\mathcal{Y}^{(1)},\ldots,\mathcal{Y}^{(q)}$, say $\mathcal{Y}^{(j)}$ for some $1\le j\le q$, contains all elements of $\mathcal{Y}^{(q+1)}$ up to $x_k+s_{q+1}$. See fig. 3. Since any of the disjoint multi-sets of size $Q$ contain just one element of $\mathcal{Y}^{(j)}$, the elements of $\mathcal{Y}^{(q+1)}$ up to $x_k+s_{q+1}$ appear in different multi-sets of size $Q=q$. We can form the disjoint multi-sets of size $q+1$ by including these common elements in the corresponding multi-sets and including the elements of $\{x_{k+1}+s_{q+1},\ldots,x_M+s_{q+1}\}$ in the remaining multi-sets arbitrarily. ∎

The condition on the channel input alphabet in the statement of theorem 2 is a sufficient condition for the channel capacity to be $\log_2 M$. However, it is not a necessary condition. For example, the statement of theorem 2 without that condition is true for the case $Q = 2$. Because in the second iteration, we do not need the arithmetic progression condition to form $M$ mutually-disjoint multi-sets of size two.

It is worth mentioning that in the proof of theorem 2, we did not use the assumption that the interference sequence is i.i.d.. In fact, the interference sequence could be any arbitrary varying sequence of the elements of $\mathcal{S}$.

The proof of theorem 2 is actually a constructive algorithm for finding $M$ (out of $M^Q$) inputs of the associated channel to be used with probability $\frac{1}{M}$ to achieve the rate $\log_2 M$ bits.

It is interesting to see that the set containing the $q$th elements of the $M$ $Q$-tuples obtained by the constructive algorithm is $\mathcal{X}$, $q = 1, \ldots, Q$. This is due to the fact that each multi-set contains one element from each $\mathcal{Y}^{(1)}, \ldots, \mathcal{Y}^{(Q)}$. Therefore, a uniform distribution on the $M$ $Q$-tuples induces uniform distributions on $X_1, \ldots, X_Q$.

## V. Uniform Transmission

In the sequel, we study the maximization of the rate $I(X_1 \cdots X_Q; Y)$ over joint pmfs $\{p_{i_1 \cdots i_Q}\}_{i_1, \ldots, i_Q = 1}^{M}$ that induce uniform marginal distributions on $X_1, \ldots, X_Q$, i.e.,

$$p_i^{(1)} = p_i^{(2)} = \cdots = p_i^{(Q)} = \frac{1}{M}, \qquad i = 1, 2, \ldots, M, \tag{14}$$

for which we show how to obtain the optimal input probability assignment. We call a transmission scheme that induces uniform distributions on $X_1, \ldots, X_Q$ as *uniform transmission*. Uniform distributions for $X_1, \ldots, X_Q$ implies uniform distribution for $X$, the input to the state-dependent channel defined in (4).

In the previous section, we established that the capacity achieving pmf for the asymptotic case of noise-free channel induces uniform distributions on $X_1, \ldots, X_Q$ (provided that we can find $M$ $Q$-tuples such that the corresponding multi-sets are mutually disjoint). Therefore, imposing the uniformity constraint given in (14) does not reduce the transmission rate in the asymptotic case of noise-free channel. However, in the general case where the noise power is not zero there will be some loss in rate due to imposing the uniformity constraint.

Imposing the uniformity constraint along with the integrality constraint (which will be explained later on in this section), however, simplifies the encoding operation for the associated channel as will be shown in this section. Furthermore, we will show in section VII that our precoding scheme with both uniformity and integrality constraints provides higher rates than the existing modulo precoding scheme of [2].

Considering the uniformity constraints in (14), the maximization of $I(X_1 \cdots X_Q; Y)$ is reduced to the linear minimization problem

$$\begin{aligned}
\min_{p_{i_1 \cdots i_Q}} \quad & \sum_{i_1=1}^{M} \cdots \sum_{i_Q=1}^{M} h_{i_1 \cdots i_Q} p_{i_1 \cdots i_Q} \\
\text{subject to} \quad & \\
& \sum_{i_2=1}^{M} \cdots \sum_{i_Q=1}^{M} p_{i_1 \cdots i_Q} = \frac{1}{M}, \qquad i_1 = 1, 2, \ldots, M, \\
& \qquad \vdots \qquad\qquad\qquad\qquad\qquad \vdots \\
& \sum_{i_1=1}^{M} \cdots \sum_{i_{Q-1}=1}^{M} p_{i_1 \cdots i_Q} = \frac{1}{M}, \qquad i_Q = 1, 2, \ldots, M, \\
& p_{i_1 \cdots i_Q} \geq 0, \qquad\qquad i_1, \ldots, i_Q = 1, 2, \ldots, M.
\end{aligned} \tag{15}$$

The equality constraints of (15) can be interpreted as the following. We assign $p_{i_1 \cdots i_Q}$ to the element $(i_1, \ldots, i_Q)$ of an $M$ by $M$ $\cdots$ by $M$ ($Q$ times) array. For $Q = 2$, the equality constraints of (15) mean that every row and every column of the array adds up to $\frac{1}{M}$. For $Q > 2$, the equality constraints can be interpreted accordingly.

The same argument used in the last part of the proof of theorem 1 can be used to show that the maximum rate with uniformity constraint is achieved by using at most $MQ - Q + 1$ inputs of the associated channel with positive probabilities. This is restated in the following corollary.

*Corollary 1:* The maximum of $I(X_1 \cdots X_Q; Y)$ over joint pmfs $\{p_{i_1 \cdots i_Q}\}_{i_1,\ldots,i_Q=1}^{M}$ that induce uniform marginal distributions on $X_1, X_2, \ldots, X_Q$ is achieved by a joint pmf with at most $MQ - Q + 1$ nonzero elements.

This result is independent of the coefficients $\{h_{i_1 \cdots i_Q}\}$. However, which probability assignment with at most $MQ - Q + 1$ nonzero elements is optimal depends on the coefficients $\{h_{i_1 \cdots i_Q}\}$. The coefficient $h_{i_1 \cdots i_Q}$ is determined by the interference levels $s_1, \ldots, s_Q$, the probability of interference levels $r_1, \ldots, r_Q$, the noise power $P_N$, and the signal points $x_1, x_2, \ldots, x_M$. The optimal probability assignment is obtained by solving the linear programming problem (15) using the simplex method [19].

### A. Two-Level Interference

If the number of interference levels is two, i.e., $Q = 2$, we can make a stronger statement than corollary 1.

*Theorem 3:* The maximum of $I(X_1 X_2; Y)$ over $\{p_{i_1 i_2}\}_{i_1,i_2=1}^{M}$ with uniform marginal pmfs for $X_1$ and $X_2$ is achieved by using exactly $M$ out of $M^2$ inputs of the associated channel with probability $\frac{1}{M}$.

*Proof:* The equality constraints of (15) can be written in matrix form as

$$\mathbf{A}\mathbf{p} = \mathbf{1}, \tag{16}$$

where $\mathbf{A}$ is a zero-one $MQ \times M^Q$ matrix, $\mathbf{p}$ is $M$ times the vector containing all $p_{i_1 \cdots i_Q}$s in lexicographical order, and $\mathbf{1}$ is the all-one $MQ \times 1$ vector.

For $Q = 2$, it is easy to check that $\mathbf{A}$ is the vertex-edge incidence matrix of $K_{M,M}$, the complete bipartite graph with $M$ vertices at each part. Therefore, $\mathbf{A}$ is a totally unimodular matrix[3] [18]. Hence, the extreme points of the feasible region $F = \{\mathbf{p} : \mathbf{A}\mathbf{p} = \mathbf{1}, \mathbf{p} \geq \mathbf{0}\}$ are integer vectors. Since the optimal value of a linear optimization problem is attained at one of the extreme points of its feasible region, the minimum in

[3] A totally unimodular matrix is a matrix for which every square submatrix has determinant 0, 1, or $-1$.

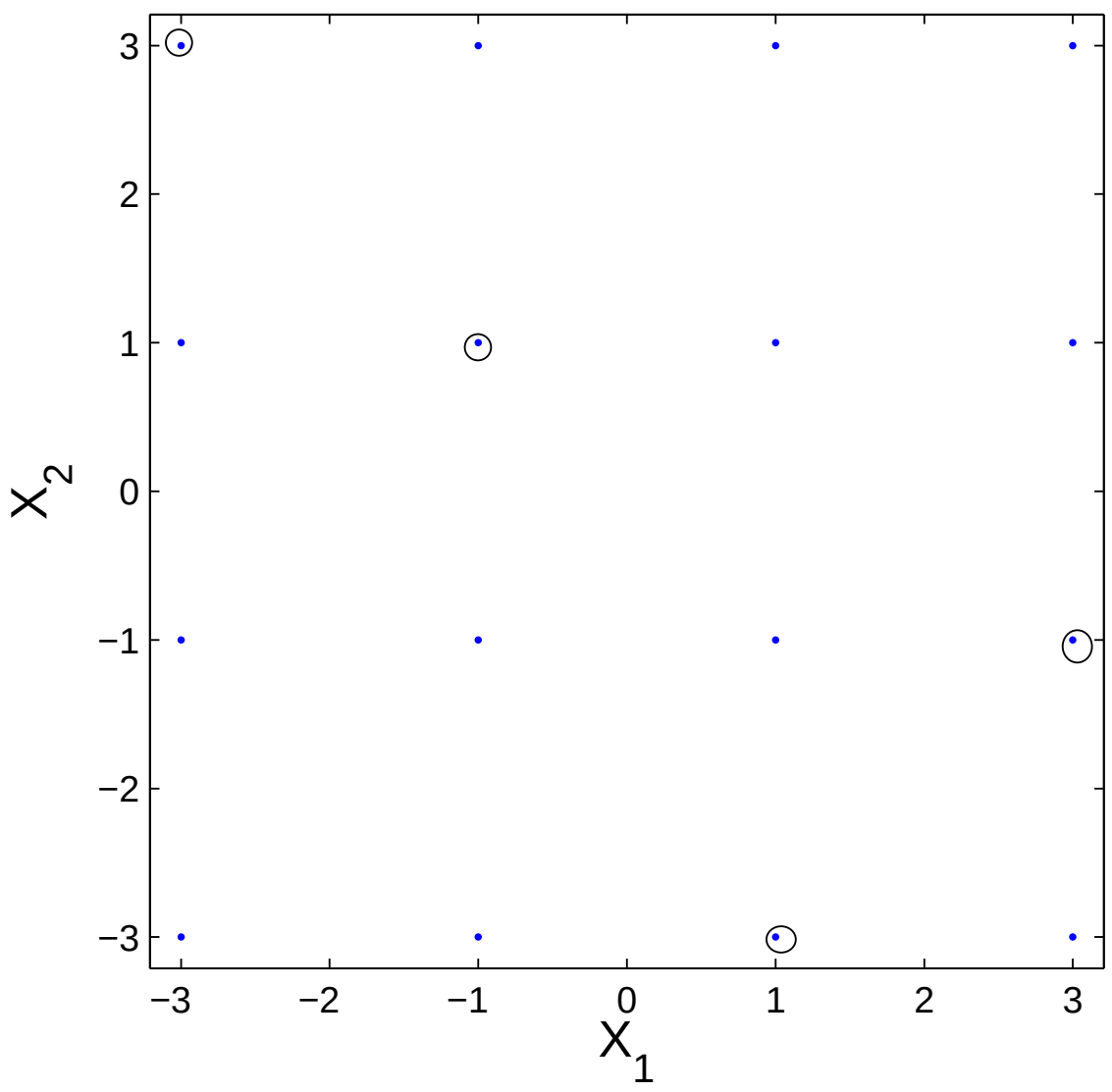

Fig. 4. Optimal solution for 4-PAM input with parameters $r_1 = r_2 = \frac{1}{2}, s_1 = -2, s_2 = +2, P_N = 1$.

(15) is achieved at an all-integer vector $\mathbf{p}^*$. Considering that $\mathbf{p}^*$ satisfies (16), it can only be a zero-one vector with exactly $M$ ones. ∎

As an example, the optimal solution for a channel with $\mathcal{X} = \{-3, -1, +1, +3\}$ and $\mathcal{S} = \{-2, 2\}$ with equiprobable interference symbols is illustrated in fig. 4. The points circled in the array correspond to the inputs to the associated channel that must be chosen with probability $\frac{1}{4}$ in order to achieve the maximum rate in the uniform transmission scenario.

Fig. 5 depicts the maximum mutual information (for the uniform transmission scenario) vs. SNR for the channel with $\mathcal{X} = \mathcal{S} = \{-1, +1\}$ and equiprobable interference symbols. The mutual information vs. SNR curve for the interference-free AWGN channel with equiprobable input alphabet $\{-1, +1\}$ is plotted for comparison purposes. As it can be seen, for low SNRs, the input probability assignment $p_{11} = p_{22} = \frac{1}{2}$ is optimal, whereas at high SNRs, the input probability assignment $p_{12} = p_{21} = \frac{1}{2}$ is optimal. The maximum achievable rate for uniform transmission is the upper envelope of the two curves corresponding to different input probability assignments. Also, it can be observed that the achievable rate approaches $\log_2 2 = 1$ bit per channel use as SNR increases complying with the fact that we established in section IV for the noise-free channel.

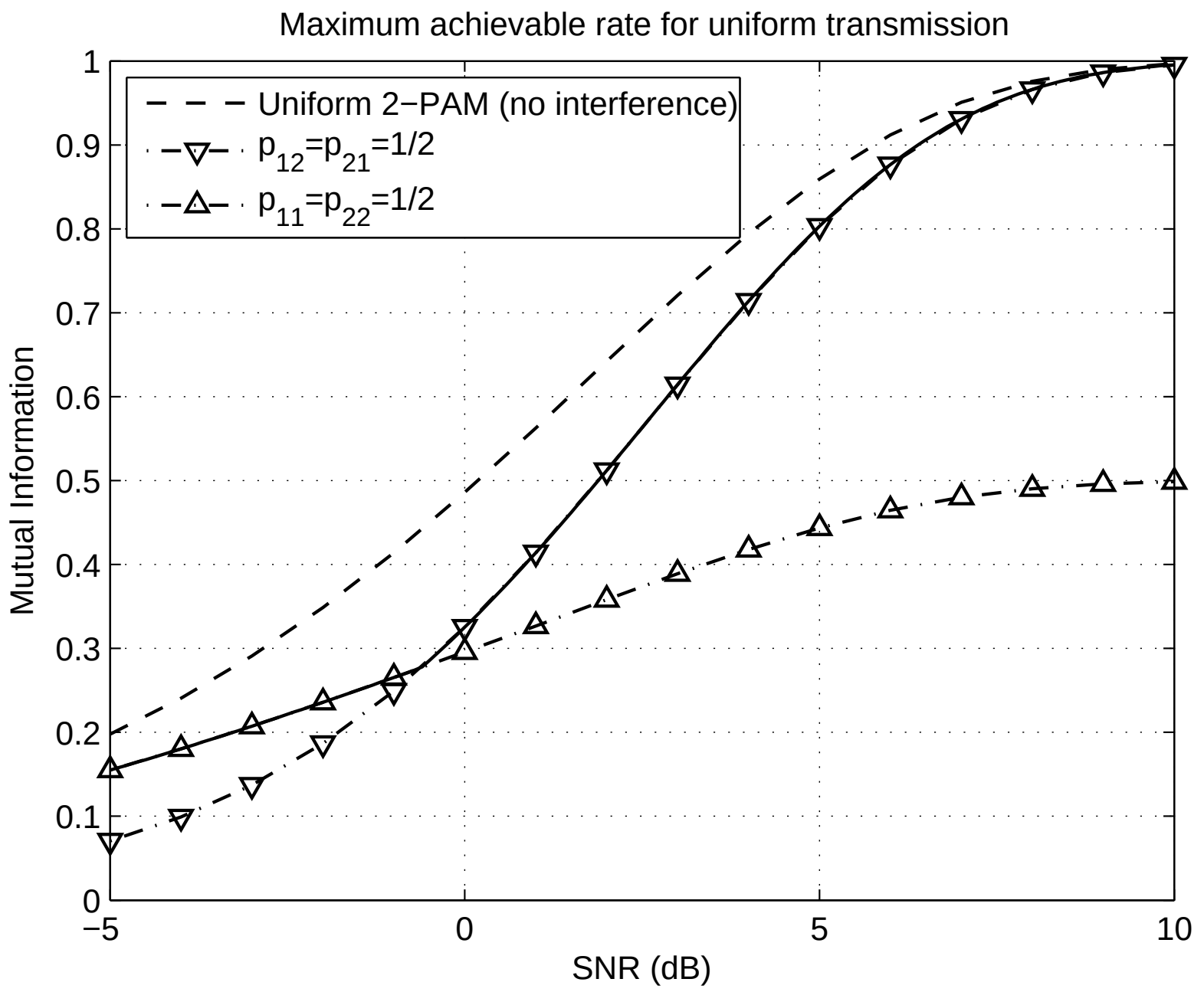

Fig. 5. Maximum mutual information vs. SNR for the channel with $\mathcal{X} = \mathcal{S} = \{-1, +1\}$ and $r_1 = r_2 = \frac{1}{2}$.

It turns out from the proof of theorem 3 that the optimum solution of the linear optimization problem, $\mathbf{p}^*$, is a zero-one vector. So, if we add the integrality constraint to the set of constraints in (16), we still obtain the same optimal solution. The resulting integer linear optimization problem is called the *assignment problem* [18], which can be solved using low-complexity algorithms such as the *Hungarian method* [19].

### *B. Integrality Constraint for the $Q$-Level Interference*

The fact that for the case $Q = 2$, there exists an optimal $\mathbf{p}$ which is a zero-one vector with exactly $M$ ones simplifies the encoding operation. Because any encoding scheme just needs to work on a subset of size $M$ of the associated channel input alphabet with equal probabilities $\frac{1}{M}$.

For $Q \neq 2$, $\mathbf{A}$ is not a totally unimodular matrix. Therefore, not all extreme points of the feasible region defined by $\mathbf{A}\mathbf{p} = \mathbf{1}, \mathbf{p} \geq \mathbf{0}$, are integer vectors. However, at the expense of possible loss in rate, we may add the integrality constraint (i.e., $\mathbf{p}$ integer) in this case. The resulting optimization problem is called the *multi-dimensional assignment problem* [20]. The optimal solution of (15) with the integrality constraint, will be a vector with exactly $M$ nonzero elements with the value $\frac{1}{M}$. Therefore, any encoding scheme just needs to use $M$ symbols of the associated channel with equal probabilities, simplifying the encoding operation.

Fig. 6 depicts the maximum mutual information for uniform transmission with the integrality constraint vs. SNR for the channel with $\mathcal{X} = \mathcal{S} = \{-3, -1, +1, +3\}$ and with equiprobable interference symbols. The mutual information vs. SNR curve for the interference-free AWGN channel with equiprobable input alphabet $\{-3, -1, +1, +3\}$ is plotted for comparison purposes. It is interesting to mention that we obtained the exact same curves as in fig. 6 without imposing the integrality constraints.

It is worth mentioning that, with the integrality constraint, the optimal solution of (15) is a joint pmf of $X_1, \ldots, X_Q$ for which $X_2, \ldots, X_Q$ can be presented as a function of $X_1$.

### C. Explicit Optimal Solutions

In the sequel, we further investigate the optimal solution of (15). It can be shown that the coefficient $h_{i_1 \cdots i_Q} = h(Y|X_1 = x_{i_1}, \ldots, X_Q = x_{i_Q})$ is a function of $x_{i_1} - x_{i_2}, x_{i_1} - x_{i_3}, \ldots, x_{i_1} - x_{i_Q}$, i.e.,

$$h_{i_1 \cdots i_Q} = g(x_{i_1} - x_{i_2}, x_{i_1} - x_{i_3}, \ldots, x_{i_1} - x_{i_Q}), \tag{17}$$

where $g$ is a given by

$$\begin{aligned} g(u_1, \ldots, u_{Q-1}) &= -\int_{-\infty}^{+\infty} \left( r_1 f_N(z) + \sum_{q=2}^{Q} r_q f_N(z + u_{q-1} + s_1 - s_q) \right) \times \\ &\qquad \log_2 \left( r_1 f_N(z) + \sum_{q=2}^{Q} r_q f_N(z + u_{q-1} + s_1 - s_q) \right) dz. \end{aligned} \tag{18}$$

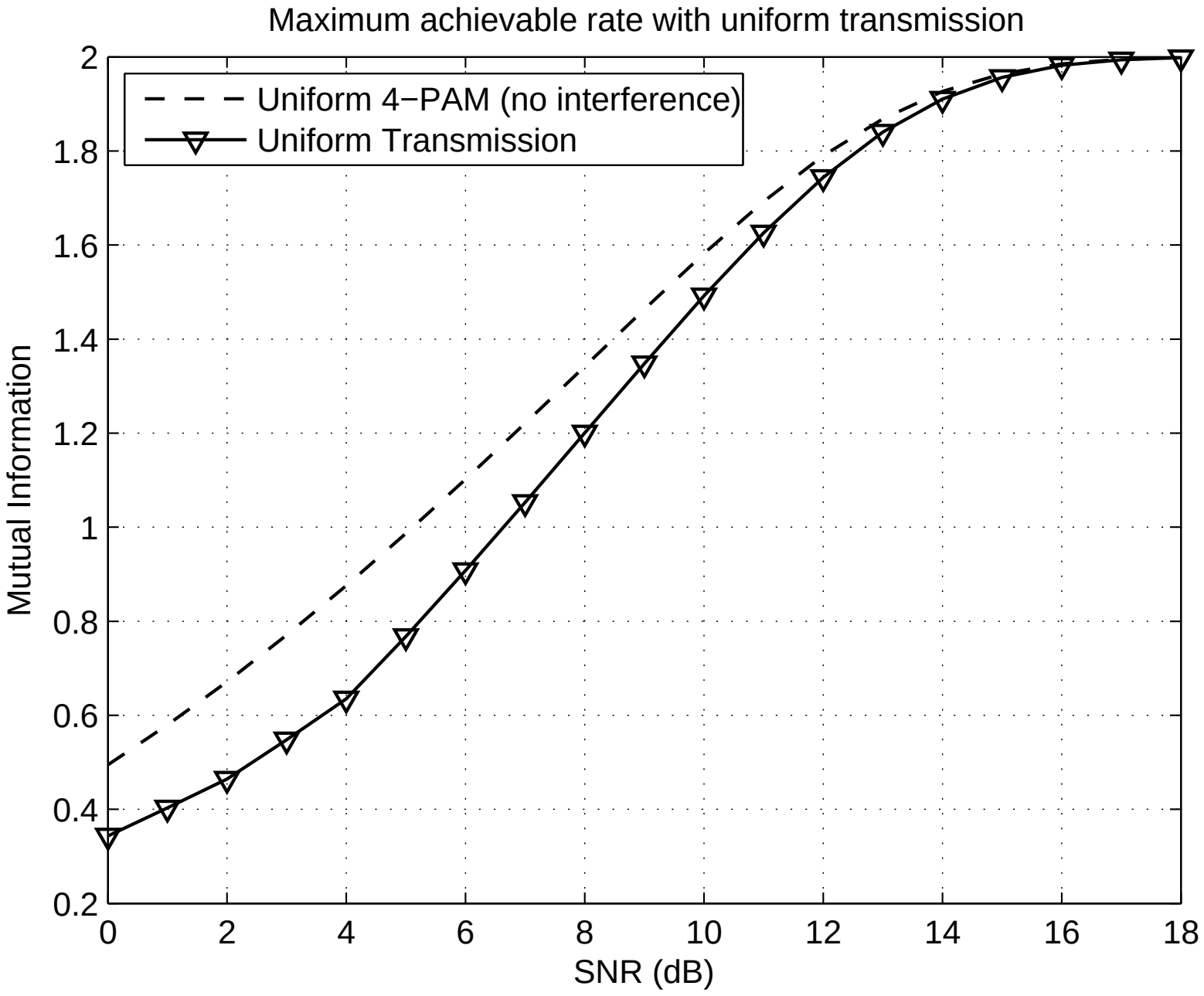

Fig. 6. Maximum mutual information vs. SNR for the channel with $\mathcal{X} = \mathcal{S} = \{-3, -1, +1, +3\}$ and $r_1 = r_2 = r_3 = r_4 = \frac{1}{4}$.

The plot of $g(.)$ for $Q = 2$ with parameters $r_1 = \frac{1}{2}, r_2 = \frac{1}{2}, s_1 = -2, s_2 = +2, P_N = 1$ is shown in fig. 7. The plot of $g(.)$ for $Q = 3$ with parameters $r_1 = r_2 = r_3 = \frac{1}{3}, s_1 = -2, s_2 = 0, s_3 = +2, P_N = 1$ is shown in fig. 8. In Appendix I, it has been shown that $g$ is lower bounded by the differential entropy of the noise, $h(N)$, and is upper-bounded by $h(N) + H(S)$, where $H(S)$ is the entropy of the discrete interference.

We may assume that $x_1$ and $x_M$ are the smallest and the largest elements of the input alphabet $\mathcal{X}$, respectively. Then the following theorem gives an explicit solution to (15) under some circumstances.

*Theorem 4:* If $g$ is convex in the $(Q-1)$-cube $\{(u_1, \ldots, u_{Q-1}) : x_1 - x_M \leq u_i \leq x_M - x_1, i = 1, 2, \ldots, Q-1\}$, then the optimal solution to (15) is

$$\tilde{p}_{i_1 \cdots i_Q} = \begin{cases} \frac{1}{M}, & \text{if} \quad i_1 = \cdots = i_Q \\ 0, & \text{otherwise.} \end{cases} \qquad (19)$$

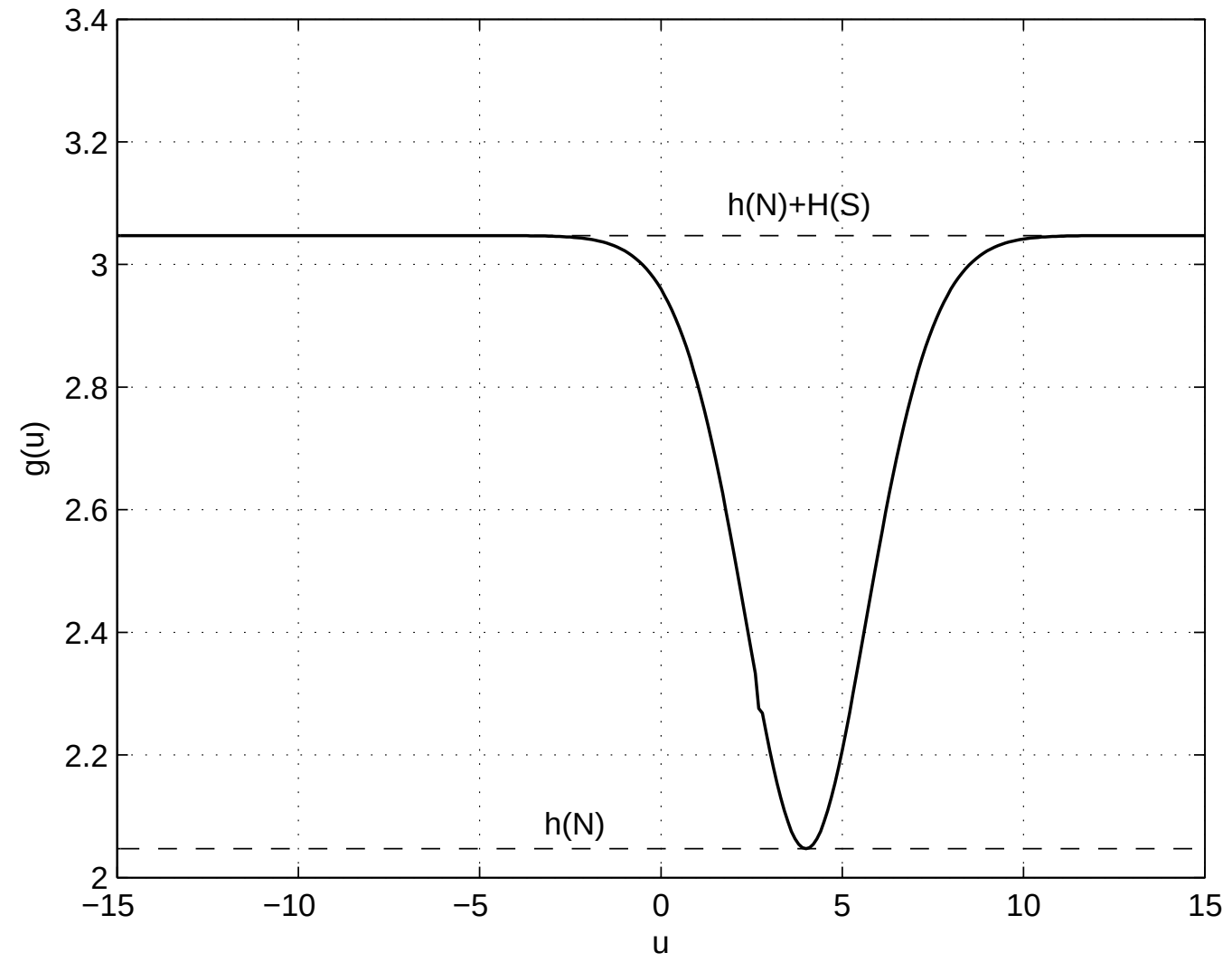

Fig. 7. The plot of $g(u)$ for $r_1 = \frac{1}{2}, r_2 = \frac{1}{2}, s_1 = -2, s_2 = +2, P_N = 1$.

*Proof:* Define random variables $U_i = X_1 - X_{i+1}$, $i = 1, \ldots, Q-1$. The objective function in (15) can be written as

$$\begin{aligned}
& \sum_{i_1=1}^{M} \cdots \sum_{i_Q=1}^{M} \Pr\left\{X_1 = x_{i_1}, \ldots, X_Q = x_{i_Q}\right\} g(x_{i_1} - x_{i_2}, \ldots, x_{i_1} - x_{i_Q}) \\
= & \sum_{j_1} \cdots \sum_{j_{Q-1}} \sum_{i_1=1}^{M} \Pr\left\{X_1 = x_{i_1}, X_2 = x_{i_1} - u_{j_1}, \ldots, X_Q = x_{i_1} - u_{j_{Q-1}}\right\} \times \\
& \qquad g(u_{j_1}, \ldots, u_{j_{Q-1}}) \\
= & \sum_{j_1} \cdots \sum_{j_{Q-1}} \sum_{i_1=1}^{M} \Pr\left\{X_1 = x_{i_1}, X_1 - X_2 = u_{j_1}, \ldots, X_1 - X_Q = u_{j_{Q-1}}\right\} \times \\
& \qquad g(u_{j_1}, \ldots, u_{j_{Q-1}}) \\
= & \sum_{j_1} \cdots \sum_{j_{Q-1}} \sum_{i_1=1}^{M} \Pr\left\{X_1 = x_{i_1}, U_1 = u_{j_1}, \ldots, U_{Q-1} = u_{j_{Q-1}}\right\} g(u_{j_1}, \ldots, u_{j_{Q-1}}) \\
= & \sum_{j_1} \cdots \sum_{j_{Q-1}} \Pr\left\{U_1 = u_{j_1}, \ldots, U_{Q-1} = u_{j_{Q-1}}\right\} g(u_{j_1}, \ldots, u_{j_{Q-1}}) \\
= & \mathrm{E}[g(U_1, \ldots, U_{Q-1})], \qquad (20)
\end{aligned}$$

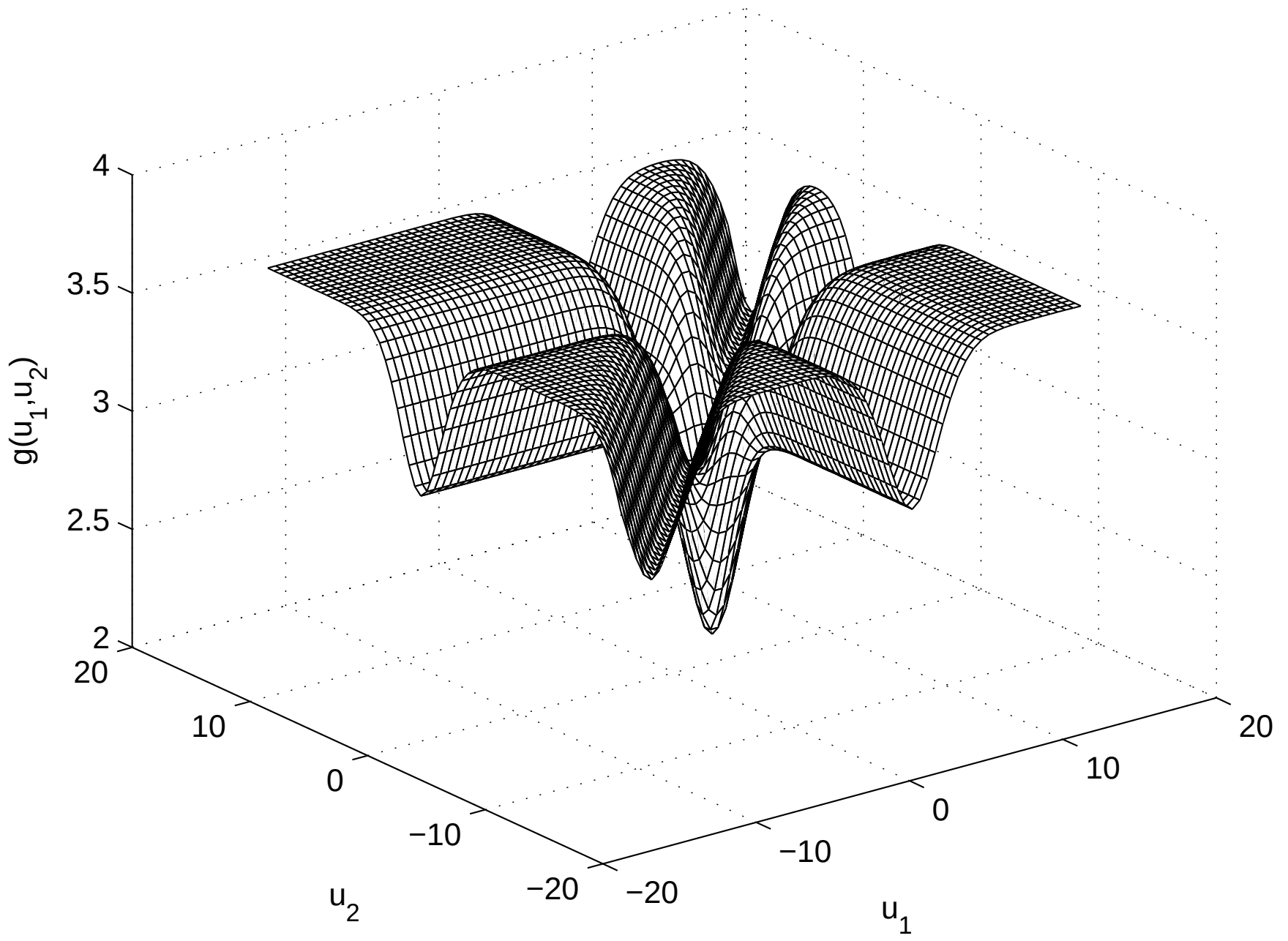

Fig. 8. The plot of $g(u_1, u_2)$ with parameters $r_1 = r_2 = r_3 = \frac{1}{3}, s_1 = -2, s_2 = 0, s_3 = +2, P_N = 1$.

where E[.] denotes the expectation operator. Now, considering the convexity of $g$, apply the Jensen's Inequality

$$\begin{aligned} \mathrm{E}[g(U_1, \ldots, U_{Q-1})] &\geq g\left(\mathrm{E}[U_1, \ldots, U_{Q-1}]\right) \\ &= g(0, \ldots, 0). \end{aligned} \tag{21}$$

Equality holds when the random variables $U_1, \ldots, U_{Q-1}$ take the value zero with probability one, or equivalently,

$$X_1 = X_2 = \cdots = X_Q. \tag{22}$$

The joint pmf in (19) satisfies both the constraints in (15) and (22), so it is the optimal solution. ■

For $Q = 2$, the convexity of $g$ in the interval $[x_1 - x_M, x_M - x_1]$ is equivalent to

$$x_M - x_1 \leq s_1 - s_2 + u^* \sqrt{P_N}, \tag{23}$$

where $u^* \approx 1.636$ and $s_1 < s_2$. The proof can be found in Appendix II. In general ($Q \geq 2$), when the power of the noise $P_N$ is sufficiently large, $g$ will be convex in the $(Q-1)$-cube.

Theorem 4 has an interesting interpretation: Given the condition of theorem 4 satisfied, the optimal precoder sends the same symbol in the channel regardless of the current interference symbol. In other words, the optimal precoder for uniform transmission ignores the interference. In fact, as it can be seen from (21), any transmission scheme that forces $X_1, \ldots, X_Q$ to have the same statistical average does not benefit from the causal knowledge of interference symbols at the transmitter if the condition of theorem 4 is satisfied. Note that this might not hold true for a capacity achieving coding scheme without any constraints on the marginal pmfs of $X_1, \ldots, X_Q$.

The following theorem holds for the case $Q = 2$ and when the input alphabet $\mathcal{X}$ is symmetric w.r.t. the origin, i.e.,

$$x_i = -x_{M+1-i}, \quad i = 1, \ldots, M. \tag{24}$$

For example, a regular PAM constellation satisfies (24).

*Theorem 5:* If the input alphabet $\mathcal{X}$ is symmetric w.r.t. the origin, and if $g$ is concave in the interval $[x_1 - x_M, x_M - x_1]$, then

$$\tilde{p}_{ij} = \begin{cases} \frac{1}{M}, & \text{if} \quad i + j = M + 1 \\ 0, & \text{otherwise.} \end{cases} \tag{25}$$

is an optimal solution to (15).

*Proof:* We rewrite (15) for the case $Q = 2$ as

$$\min_{p_{ij}} \quad \sum_{i=1}^{M}\sum_{j=1}^{M} h_{ij}p_{ij}$$

subject to

$$\begin{aligned}
&\sum_{j=1}^{M} p_{ij} = \frac{1}{M}, \quad i = 1, 2, \ldots, M, \\
&\sum_{i=1}^{M} p_{ij} = \frac{1}{M}, \quad j = 1, 2, \ldots, M, \\
&p_{ij} \geq 0, \qquad i, j = 1, 2, \ldots, M.
\end{aligned} \tag{26}$$

We assign $p_{ij}$ to the element $(i, j)$ of an $M$ by $M$ array (See fig. 4). The equality constraints of (26) mean that every row and every column of the array adds up to $\frac{1}{M}$. We make the observation that if $\{p_{ij}\}_{i,j=1,2,\ldots,M}$ is a feasible solution of (26), then $\{q_{ij}\}_{i,j=1,2,\ldots,M}$, where $q_{ij} = p_{(M+1-j)(M+1-i)}$, will also be a feasible solution of (26). Furthermore, due to (24) and the fact that $h_{ij} = g(x_i - x_j)$, $\{p_{ij}\}$ and $\{q_{ij}\}$ yield the same objective value. Therefore, if $\{p_{ij}\}$ is an optimal solution of (26), $\{q_{ij}\}$ will be an optimal solution too. The convex combination of the two optimal solutions $\{\theta_{ij} = \frac{1}{2}p_{ij} + \frac{1}{2}q_{ij}\}$ is also an optimal solution with the following symmetry property

$$\theta_{ij} = \theta_{(M+1-j)(M+1-i)}. \tag{27}$$

In fact, (27) describes a solution which is symmetric w.r.t. the main diagonal of the array. So far, we have established the existence of an optimal solution to (26) with the symmetry property (27). Now, suppose that a symmetric optimal solution to (26) has nonzero entries

$$p_{ij} = p_{(M+1-j)(M+1-i)} = p, \tag{28}$$

where $i + j \neq M + 1$. Now, if we add $p$ to the main diagonal entries $p_{(M+1-j)j}$ and $p_{i(M+1-i)}$ and turn $p_{ij}$ and $p_{(M+1-j)(M+1-i)}$ to zero, the constraints of (26) are not violated.

However, the change in the objective function will be proportional to

$$h(Y|X_1 = x_i, X_2 = x_{M+1-i}) + h(Y|X_1 = x_{M+1-j}, X_2 = x_j)$$
$$-h(Y|X_1 = x_i, X_2 = x_j) - h(Y|X_1 = x_{M+1-j}, X_2 = x_{M+1-i}),$$

which is equal to $g(2x_i)+g(-2x_j)-2g(x_i-x_j)$ which is non-positive by concavity of $g$. Hence, we have not increased the objective value by the process described above. We can repeat the process until all nonzero entries lie on the main diagonal without increasing the objective value. Therefore, (25) is an optimal solution of (26). ■

It can be shown that $g$ is concave in the interval $[x_1 - x_M, x_M - x_1]$ if and only if

$$x_M - x_1 \leq s_2 - s_1 - u_0\sqrt{P_N}. \tag{29}$$

See Appendix II for the proof.

## VI. Optimal Precoding

The general structure of a communication system for the channel defined in (4) is shown in fig. 9. In fact, fig. 9 is the same as fig. 2 for the special case of the state-dependent channel defined in (4). Any encoding and decoding scheme for the associated channel can be translated to an encoding and decoding scheme for the original channel defined in (4). A message $w$ is encoded to a block of length $n$ composed of input symbols of the associated channel $t \sim (x_{i_1}, x_{i_2}, \ldots, x_{i_Q})$. There are $M^Q$ input symbols. However, we showed that the maximum rate with uniformity and integrality constraints can be achieved by using just $M$ input symbols of the associated channel with equal probabilities. The optimal $M$ input symbols of the associated channel are obtained by solving the linear programming problem (15) with the integrality constraint. Those $M$ input symbols of the associated channel define the optimal precoding operation: For any $t$ that belongs to the set of $M$ optimal input symbols, the precoder sends the $q$th component of $t$ if the current interference symbol is $s_q$, $q = 1, \ldots, Q$. Based on the received sequence, the receiver decodes $\hat{w}$ as the transmitted message.

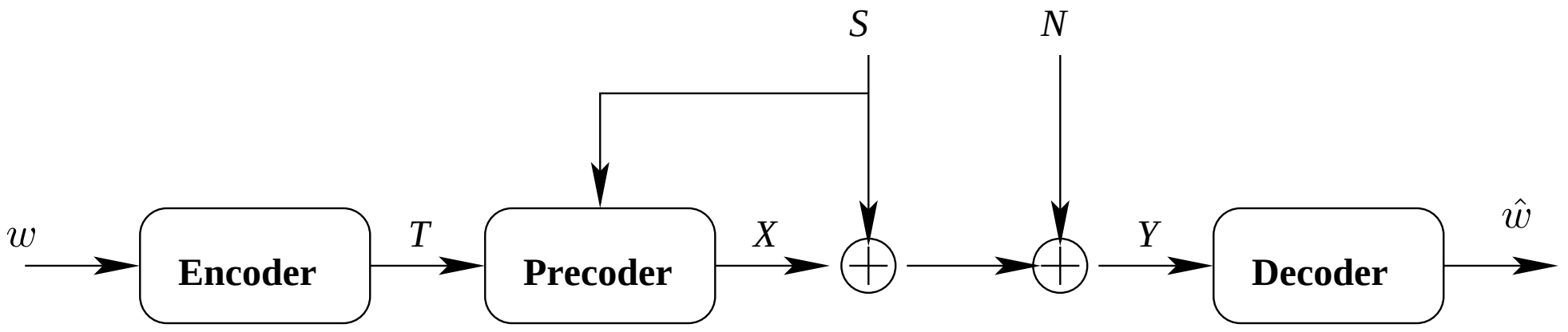

Fig. 9. General structure of the communication system for channels with causally-known discrete interference.

## VII. Extension to Continuous Input Alphabet

We can extend the uniform transmission scheme introduced in section V to the case where the channel input alphabet $\mathcal{X}$ is continuous. For the continuous input alphabet case, we consider the maximization of the transmission rate $I(X_1 \cdots X_Q; Y)$ over joint pdfs $f_{X_1 \cdots X_Q}(x_1, \ldots, x_Q)$ that induce uniform marginal distributions on $X_1, \ldots, X_Q$ in the interval $A_\Delta = \left[-\frac{\Delta}{2}, \frac{\Delta}{2}\right]$.

Since $h(Y)$ is the same for all joint pdfs $f_{X_1 \cdots X_Q}(x_1, \ldots, x_Q)$ that induce uniform marginal pdfs on $X_1, \ldots, X_Q$, the maximization of the transmission rate reduces to the linear minimization problem

$$\begin{aligned}
\min_{f_{X_1 \cdots X_Q}} \quad & \int_{-\frac{\Delta}{2}}^{\frac{\Delta}{2}} \cdots \int_{-\frac{\Delta}{2}}^{\frac{\Delta}{2}} h(x_1, \ldots, x_Q) f_{X_1 \cdots X_Q}(x_1, \ldots, x_Q) dx_1 \cdots dx_Q \\
\text{subject to} \quad & \\
& \int_{-\frac{\Delta}{2}}^{\frac{\Delta}{2}} \cdots \int_{-\frac{\Delta}{2}}^{\frac{\Delta}{2}} f_{X_1 \cdots X_Q}(x_1, \ldots, x_Q) dx_2 \cdots dx_Q = \frac{1}{\Delta}, \qquad x_1 \in A_\Delta, \\
& \vdots \qquad\qquad\qquad\qquad \vdots \qquad\qquad \vdots \\
& \int_{-\frac{\Delta}{2}}^{\frac{\Delta}{2}} \cdots \int_{-\frac{\Delta}{2}}^{\frac{\Delta}{2}} f_{X_1 \cdots X_Q}(x_1, \ldots, x_Q) dx_1 \cdots dx_{Q-1} = \frac{1}{\Delta}, \qquad x_Q \in A_\Delta, \\
& f_{X_1 \cdots X_Q}(x_1, \ldots, x_Q) \geq 0, \qquad x_1, \ldots, x_Q \in A_\Delta,
\end{aligned} \tag{30}$$

where $h(x_1, \ldots, x_Q) = h(Y | X_1 = x_1, \ldots, X_Q = x_Q)$. We are interested in solutions to (30) that are of the form

$$f_{X_1 \cdots X_Q}(x_1, \ldots, x_Q) = \frac{1}{\Delta} \delta \left( |x_2 - \xi_1(x_1)| + |x_3 - \xi_2(x_1)| + \cdots + |x_Q - \xi_{Q-1}(x_1)| \right), \tag{31}$$

where $\delta(.)$ is the Dirac's delta function, $|.|$ denote absolute value, and $\xi_1, \xi_2, \ldots, \xi_{Q-1}$ are bijective functions from $A_\Delta$ to $A_\Delta$.

The joint pdf in (31) describes random variables $X_1, \ldots, X_Q$, $Q-1$ of which are functions of the other random variable. Solutions of the form (31) can be considered as the continuous extension of solutions to (15) with the integrality constraint for the discrete input alphabet case. It is easy to check that (31), with the given condition that $\xi_1, \xi_2, \ldots, \xi_{Q-1}$ are bijective function from $A_\Delta$ to $A_\Delta$, satisfies the constraints in (30). The objective value corresponding to the joint pdf (31) is

$$\frac{1}{\Delta}\int_{-\frac{\Delta}{2}}^{\frac{\Delta}{2}} h\left(x_1, \xi_1(x_1), \ldots, \xi_{Q-1}(x_1)\right) dx_1, \tag{32}$$

which is to be minimized over bijective functions $\xi_1, \xi_2, \ldots, \xi_{Q-1}$.

### A. Comparison to Modulo Precoding

The modulo precoding was originally proposed by Tomlinson and Harashima [21], [22] for the ISI channel. Then it was extended in [2] as a precoding method for channels with known (discrete or continuous) interference at the transmitter. The main idea is as follows. Based on the input symbol of the associated channel $V$ and the current interference symbol $S$, the precoder sends [2]

$$X = [V - \alpha S] \mod \Delta, \tag{33}$$

where $\alpha = \frac{P_X}{P_X + P_N}$ ($P_X$ is the power of $X$) and $V$ is distributed uniformly in $A_\Delta$.

In our setting where the interference is discrete with $Q$ levels, (33) results in

$$X_q = [V - \alpha s_q] \mod \Delta, \qquad q = 1, \ldots, Q, \tag{34}$$

where $X_q$ is the random variable that represents the channel input when the current interference symbol is $s_q$, $q = 1, \ldots, Q$. Since $V$ is uniformly distributed in $A_\Delta$, $X_1, \ldots, X_Q$ will be uniformly distributed in $A_\Delta$. Therefore, modulo precoding is indeed a uniform transmission scheme. We can remove $V$ from the above equations and express $X_2, \ldots, X_Q$ in terms of $X_1$ as

$$X_q = [X_1 + \alpha(s_1 - s_q)] \mod \Delta, \qquad q = 2, \ldots, Q. \tag{35}$$

Since $X_2, \ldots, X_Q$ are functions of $X_1$, the joint pdf $f_{X_1 \cdots X_Q}(x_1, \ldots, x_Q)$ corresponding to the modulo precoding fits in the category of joint pdfs in (31). The bijective functions corresponding to the modulo precoding are given by (35). These functions are circular shifts of each other.

The modulo precoding corresponds to a feasible solution to (30) which is not an optimal solution. For example, we may follow the line of proof of theorem 4 to show that for large $P_N$, where $g$ becomes convex in the hyper-cube $\{(u_1, \ldots, u_{Q-1}) : -\Delta \leq u_i \leq \Delta, i = 1, \ldots, Q-1\}$, the optimal bijective functions are given by $\xi_1(x) = \cdots = \xi_{Q-1}(x) = x$, which are different from the functions given in (35).

To make the example more specific, consider a channel with $\mathcal{X} = A_\Delta = [-1, +1]$ and $\mathcal{S} = \{-\frac{1}{2}, +\frac{1}{2}\}$. According to (23), $g(u)$ will be convex if we choose $P_N = 3.363$. Then we will have $\alpha = \frac{P_X}{P_X + P_N} = \frac{0.333}{0.333+3.363} \approx 0.09$. Therefore, the bijective function corresponding to modulo precoding is given by

$$X_2 = [X_1 - 0.09] \mod 2, \tag{36}$$

while the optimal precoding corresponds to $X_2 = X_1$ in this example.

## VIII. Conclusion

In this paper, we investigated $M$-ary signal transmission over AWGN channel with additive $Q$-level interference, where the sequence of i.i.d. interference symbols is known causally at the transmitter. According to Shannon's theorem for channels with side information at the transmitter, the capacity of our channel is the same as the capacity of an associated regular (without state) channel with $M^Q$ input symbols. We proved that by using at most $MQ - Q + 1$ (out of $M^Q$) input symbols the capacity is achievable.

For the noise-free channel, provided that the signal points are equally spaced, we proposed a one-shot coding scheme that uses $M$ input symbols of the associated channel to achieves the capacity $\log_2 M$ bits regardless of the interference.

We considered the maximization of the transmission rate with the constraint that $X_1, \ldots, X_Q$ are uniformly distributed over the channel input alphabet. For this so called uniform transmission, the optimal input probability assignment (again with at most $MQ - Q + 1$ nonzero elements) can be obtained by solving the linear optimization problem (15). The optimal solution to (15) with the integrality constraint has exactly $M$ nonzero elements. For the case $Q = 2$, we showed that the integrality constraint does not reduce the maximum achievable rate. The loss in rate (if there is any) by imposing the integrality constraint for the general case is a problem to be explored.

## Appendix I
## Bounds For $h(Y|X_1 = x_{i_1}, \ldots, X_Q = x_{i_Q})$

Denote by $\tilde{S}$ the random variable that takes on $x_{i_1} + s_1, x_{i_2} + s_2, \ldots, x_{i_Q} + s_Q$ with probabilities $r_1, r_2, \ldots, r_Q$, respectively. Also, denote by $\tilde{Y}$ the random variable $Y|X_1 = x_{i_1}, \ldots, X_Q = x_{i_Q}$. Then

$$\tilde{Y} = \tilde{S} + N. \tag{37}$$

Since

$$0 \leq I(\tilde{Y}; \tilde{S}) \leq H(\tilde{S}), \tag{38}$$

we have

$$0 \leq h(\tilde{Y}) - h(\tilde{Y}|\tilde{S}) \leq H(\tilde{S}), \tag{39}$$

or equivalently,

$$\begin{aligned} h(N) \leq h(\tilde{Y}) \quad &\leq \quad h(N) + H(\tilde{S}) \\ &= \quad h(N) + H(S). \end{aligned} \tag{40}$$

## APPENDIX II

### NECESSARY AND SUFFICIENT CONDITIONS FOR THE CONVEXITY/CONCAVITY OF $g$

The function $g$ given in (18) for the case $Q = 2$ can be considered as a function of $u$ and parameters $s_1, s_2, P_N$ as

$$\begin{aligned} g(u) &= g(u, s_1, s_2, P_N) \\ &= g(u + s_1 - s_2, 0, 0, P_N) \\ &= g\left(\frac{u + s_1 - s_2}{\sqrt{P_N}}, 0, 0, 1\right) + \log_2 \sqrt{P_N}. \end{aligned} \tag{41}$$

Denote by $u_0$ and $-u_0$ the inflection points of $g(u, 0, 0, 1)$. We can obtain $u_0$ numerically as $u_0 \approxeq 1.636$. Then the inflection points of $g(u)$ are

$$\alpha_1 = s_2 - s_1 - u_0\sqrt{P_N}, \tag{42}$$

$$\alpha_2 = s_2 - s_1 + u_0\sqrt{P_N}, \tag{43}$$

The function $g$ is convex in the interval $[\alpha_1, \alpha_2]$ and is concave anywhere else.

The function $g$ is convex in the interval $[x_1 - x_M, x_M - x_1]$ if and only if $[x_1 - x_M, x_M - x_1] \subseteq [\alpha_1, \alpha_2]$. This gives (23).

The function $g$ is concave in the interval $[x_1 - x_M, x_M - x_1]$ if and only if $[x_1 - x_M, x_M - x_1] \subseteq (-\infty, \alpha_1]$ or $[x_1 - x_M, x_M - x_1] \subseteq [\alpha_2, \infty)$. This gives (29).

## REFERENCES

[1] M. H. M. Costa, "Writing on dirty paper," *IEEE Trans. Inform. Theory*, vol. 29, no. 3, pp. 439-441, May 1983.

[2] U. Erez, S. Shamai, and R. Zamir, "Capacity and lattice strategies for canceling known interference," *IEEE Trans. Inform. Theory*, vol. 51, no. 11, pp. 3820-3833, Nov. 2005.

[3] G. Caire and S. Shamai,"On achievable throughput of a multiple antenna Gaussian broadcast channel," *IEEE Trans. Inform. Theory*, vol. 49, no. 7, pp. 1691-1706, Jul. 2003.

[4] W. Yu and J. M. Cioffi,"Sum capacity of Gaussian vector broadcast channels," *IEEE Trans. Inform. Theory*, vol. 50, no. 9, pp. 1875-1892, Sep. 2004.

[5] S. Viswanath, N. Jindal, and A. Goldsmith,"Duality, achievable rates, and sum-rate capacity of Gaussian MIMO broadcsat channels," *IEEE Trans. Inform. Theory*, vol. 49, no. 10, pp. 2658-2668, Oct. 2003.

[6] P. Viswanath and D. Tse,“Sum capacity of the multiple-antenna Gaussian broadcast channel and uplink-downlink duality,” *IEEE Trans. Inform. Theory*, vol. 49, no. 7, pp. 1912-1921, Jul. 2003.

[7] H. Weingarten, Yosef Steinberg, and S. Shamai, “The capacity region of Gaussian multiple-input multiple-output channel,” *IEEE Trans. Inform. Theory*, vol. 52, no. 9, pp. 3936-3964, Sept. 2006

[8] B. Chen and G. W. Wornell, “Quantization index modulation: A class of provably good methods for digital watermarking and information embedding,” *IEEE Trans. Inform. Theory*, vol. 47, no. 4, pp. 1423-1443, May 2001.

[9] A. Cohen and A. Lapidoth,“The Gaussian watermarking game,” *IEEE Trans. Inform. Theory*, vol. 48, no. 6, pp. 1639-1667, Jun. 2002.

[10] P. Moulin and J. A. O’Sullivan,“Information-Theoretic Analysis of Information Hiding,” *IEEE Trans. Inform. Theory*, vol. 49, no. 3, pp. 563-593, Mar. 2003.

[11] C. E. Shannon, “Channels with side information at the transmitter,” *IBM Journal of Research and Development*, vol. 2, pp. 289-293, Oct. 1958.

[12] A. V. Kuznetsov and B. S. Tsybakov, “Coding in a memory with defective cells,” *Probl. Pered. Inform.*, vol. 10, no. 2, pp. 52-60, Apr.-June 1974.

[13] S. Gel’fand and M. Pinsker, “Coding for channel with random parameters,” *Problems of Control and Information Theory*, vol. 9, no. 1, pp. 19-31, Jan. 1980.

[14] M. Salehi, “Capacity and coding for memories with real-time noisy defect information at the encoder and decoder,” *Proc. Inst. Elec. Eng.-Pt. I*, vol. 139, no. 2, pp. 113-117, Apr. 1992.

[15] G. caire and S. Shamai,“On the capacity of some channels with channel state information,” *IEEE Trans. Inform. Theory*, vol. 45, no. 6, pp. 2007-2019, Sep. 1999.

[16] C. Heegard and A. El Gamal,“On the capacity of computer memories with defects,” *IEEE Trans. Inform. Theory*, vol. 29, no. 5, pp. 731-739, Sep. 1983.

[17] A. Rosenzweig, Y. Steinberg, and S. Shamai,“On channels with partial state information at the transmitter,” *IEEE Trans. Inform. Theory*, vol. 51, no. 5, pp. 1817-1830, May 2005.

[18] G. Nemhauser and L. Wolsey, *Integer and combinatorial optimization*, John Wiley & Sons, 1988.

[19] B. Krekó, *Linear Programming*, Translated by J. H. L. Ahrens and C. M. Safe. Sir Isaac Pitman & Sons Ltd., 1968.

[20] W. P. Pierskalla, “The multidimensional assignment problem,” *Operations Research* 16, p. 422-431, 1968.

[21] M. Tomlinson, “New automatic equalizer employing modulo arithmetic,” *Electron. Lett.*, vol. 7, pp. 138-139, Mar. 1971.

[22] M. Miyakawa and H. Harashima, “A method of code conversion for a digital communication channel with intersymbol interference,” *Trans. Inst. Electron. Commun. Eng. Japan*, vol. 52-A, pp. 272-273, Jun. 1969.